\documentclass[%
 reprint,
superscriptaddress,
showpacs,showkeys,preprintnumbers,
 amsmath,amssymb,
 aps,
]{revtex4-1}

\usepackage{graphicx}
\usepackage{dcolumn}
\usepackage{bm}
\usepackage{here}
\usepackage[dvips]{color}
\usepackage{multirow}
\usepackage[normalem]{ulem}

\makeatletter
  \def\erf{\mathop{\operator@font erf}\nolimits}
  \def\erfc{\mathop{\operator@font erfc}\nolimits}
  \def\Erf{\mathop{\operator@font Erf}\nolimits}
  \def\Shi{\mathop{\operator@font Shi}\nolimits}
  \def\Chi{\mathop{\operator@font Chi}\nolimits}
  \def\Ei{\mathop{\operator@font Ei}\nolimits}
  \def\cosec{\mathop{\operator@font cosec}\nolimits}
  \def\sech{\mathop{\operator@font sech}\nolimits}
  \def\cosech{\mathop{\operator@font cosech}\nolimits}
  \newcommand\hypgeo[2]{{}_{#1}{\operator@font F}_{#2}}
  \def\Re{\mathop{\operator@font Re}\nolimits}
  \def\Im{\mathop{\operator@font Im}\nolimits}
\makeatother

\begin{document}


\title{High sensitivity of a future search for P-odd/T-odd interactions on the 0.75~eV $p$-wave resonance in $\vec{n}+^{139}\vec{\rm La}$ forward transmission determined using pulsed neutron beam} 


\def\affNagoya{Nagoya University, Furocho, Chikusa, Nagoya 464-8602, Japan}
\def\affKyushu{Kyushu University, 744 Motooka, Nishi, Fukuoka 819-0395, Japan}
\def\affJAEA{Japan Atomic Energy Agency, 2-4 Shirakata, Tokai, Ibaraki 319-1195, Japan}
\def\affTokyoTech{Tokyo Institute of Technology, Meguro, Tokyo 152-8551, Japan}
\def\affIbaraki{Ibaraki University, 2-1-1 Bunkyo, Mito, Ibaraki 310-8512. Japan}
\def\affRCNP{Osaka University, Ibaraki, Osaka 567-0047, Japan}
\def\affIndiana{Indiana University, Bloomington, Indiana 47401, USA}
\def\affLosAlamos{Los Alamos National Laboratory, Los Alamos, NM 87545, USA}
\def\affKEK{High Energy Accelerator Research Organization,1-1 Oho, Tsukuba, Ibaraki 305-0801, Japan}
\def\affTohoku{Tohoku University, 2-1-1 Katahira, Aoba, Sendai, 980-8576 Japan}
\def\affCross{Comprehensive Research Organization for Science and Society, Tokai, Ibaraki 319-1106, Japan}
\def\affSouthCarolina{University of South Carolina, Columbia, South Carolina 29208, USA}

\author{R.~Nakabe}
\affiliation{\affNagoya}
\author{C.~J.~Auton}
\affiliation{\affIndiana}
\affiliation{\affNagoya}
\author{S.~Endo}
\affiliation{\affJAEA}
\affiliation{\affNagoya}
\author{H.~Fujioka}
\affiliation{\affTokyoTech}
\author{V.~Gudkov}
\affiliation{\affSouthCarolina}
\author{K.~Hirota}
\affiliation{\affKEK}
\author{I.~Ide}
\affiliation{\affNagoya}
\author{T.~Ino}
\affiliation{\affKEK}
\author{M.~Ishikado}
\affiliation{\affCross}
\author{W.~Kambara}
\affiliation{\affJAEA}
\author{S.~Kawamura}
\affiliation{\affNagoya}
\affiliation{\affJAEA}
\author{A.~Kimura}
\affiliation{\affJAEA}
\author{M.~Kitaguchi}
\affiliation{\affNagoya}
\author{R.~Kobayashi}
\affiliation{\affIbaraki}
\author{T.~Okamura}
\affiliation{\affKEK}
\author{T.~Oku}
\affiliation{\affJAEA}
\affiliation{\affIbaraki}
\author{T.~Okudaira}
\affiliation{\affNagoya}
\affiliation{\affJAEA}
\author{M.~Okuizumi}
\affiliation{\affNagoya}
\author{J.~G.~Otero~Munoz}
\affiliation{\affIndiana}
\author{J.~D.~Parker}
\affiliation{\affCross}
\author{K.~Sakai}
\affiliation{\affJAEA}
\author{T.~Shima}
\affiliation{\affRCNP}
\author{H.~M.~Shimizu}
\affiliation{\affNagoya}
\author{T.~Shinohara}
\affiliation{\affJAEA}
\author{W.~M.~Snow}
\affiliation{\affIndiana}
\author{S.~Takada}
\affiliation{\affTohoku}
\affiliation{\affJAEA}
\author{R.~Takahashi}
\affiliation{\affJAEA}
\author{S.~Takahashi}
\affiliation{\affIbaraki}
\affiliation{\affJAEA}
\author{Y.~Tsuchikawa}
\affiliation{\affJAEA}
\author{T.~Yoshioka}
\affiliation{\affKyushu}


\date{\today}


\begin{abstract}
Neutron transmission experiments can offer a new type of highly sensitive search for time-reversal invariance violating (TRIV) effects in nucleon-nucleon interactions via the same enhancement mechanism observed for large parity violating (PV) effects in neutron-induced compound nuclear processes.
In these compound processes, the TRIV cross-section is given as the product of the PV cross-section, a spin-factor $\kappa$, and a ratio of TRIV and PV matrix elements.
We determined $\kappa$ to be $0.59\pm0.05$ for $^{139}$La+$n$ using both $(n, \gamma)$ spectroscopy and ($\vec{n}+^{139}\vec{\rm La}$) transmission. This result quantifies for the first time the high sensitivity of the $^{139}$La 0.75~eV $p$-wave resonance in a future search for P-odd/T-odd interactions in ($\vec{n}+^{139}\vec{\rm La}$) forward transmission.
\end{abstract}

\keywords{neutron induced compound nuclei,
polarized epithermal neutrons, 
polarized nuclear target}
\maketitle


The search for time-reversal invariance violating (TRIV) effects is one of many efficient methods to search for new physics beyond the present standard model of elementary particles.
It has been suggested that compound nuclear processes may provide a highly sensitive search for TRIV effects in the nucleon-nucleon interaction on the basis of the experimental fact that extremely large parity violating (PV) effects are observed in $p$-wave compound resonances from the small contribution of the weak interaction in the nuclear interaction~\cite{Bunakov1982,BUNAKOV198393,Gudkov1992,Bowman2014}.
Previous experiments have reported that the PV effect measured by the helicity dependence of the neutron absorption reaction of $^{139}$La is approximately 10\% of the 0.75~eV $p$-wave resonance cross-section~\cite{alf83,MASUDA1989,LANL91,shi93}, corresponding to 2\% of the total neutron cross-section. This is exceedingly larger than that of the nucleon-nucleon interaction, which has been measured to be on the order of $10^{-8}-10^{-7}$~\cite{Kistryn1987,npdgamma18}. PV and TRIV effects are both understood to be a result of the non-vanishing statistical variance of a large number of perturbative contributions due to a large number of degrees of freedom in the compound nuclear process~\cite{Sushkov1982,BUNAKOV198393}.
These TRIV effects can be related to the PV effects at the same $p$-wave resonances through a spin-dependent factor, quantified by the spin of the compound nucleus and the $p$-wave partial neutron widths~\cite{BUNAKOV198393,Gudkov1990}. 
The TRIV and PV cross-sections $\Delta \sigma_{\rm \not{T}\not{P}}$ and $\Delta \sigma_{\rm \not{P}}$ can be related to~\cite{Gudkov1990}
\begin{equation}
\Delta \sigma_{\rm \not{T}\not{P}}=\kappa\frac{W_{\rm T}}{W}\Delta \sigma_{\rm \not{P}}
\label{eq_sigmaT}
\end{equation}
where $W_{\rm T}$ and $W$ are the TRIV and PV matrix elements~\cite{Bunakov1982}. The variable $\kappa$ is the spin-dependent factor for the compound nucleus spin $J=I+1/2$ expressed as \cite{Gudkov2018}
\begin{equation}
    \kappa = \frac{I}{I+1}\left(1+\frac{1}{2}\sqrt{\frac{2I+3}{I}}\frac{y}{x}\right)
 \label{eq_kappa}
\end{equation}
where $I$ is the target nucleus spin. The variables $x$ and $y$ are the ratios of $p$-wave partial widths for $j=1/2$ and $j=3/2$ to the $p$-wave total neutron width $\Gamma_p^n$, where $\vec{j}=\vec{l}+\vec{s}$ with $l$ and $s$ as the neutron orbital angular momentum and its spin. They are defined as $x^2=\Gamma_{p,j=1/2}^n/\Gamma_p^n$ and $y^2=\Gamma_{p,j=3/2}^n/\Gamma_p^n$~\cite{fla85}, which satisfies the constraint, $x^2+y^2=1$, due to $\Gamma_p^n=\Gamma_{p,j=1/2}^n+\Gamma_{p,j=3/2}^n$. Hence, the corresponding mixing angle $\phi$ can be defined as $x=\cos\phi$ and $y=\sin\phi$.

The four possible solutions of $x$ and $y$ were obtained from the results of the spin-dependent cross-section at the 0.75~eV $p$-wave resonance with polarized neutron transmission through a transversely polarized $^{139}$La target as reported in Ref.~\cite{okudaira2023_ImB}. The total neutron cross-section of the $p$-wave resonance can be described by the spin-independent cross-section $\sigma_{\rm 0}$ and spin-dependent cross-section $\sigma_{\rm S}$ as $\sigma = \sigma_{\rm 0}\pm\sigma_{\rm S}(\vec{s}\cdot\vec{I})$, where $\vec{s}$ and $\vec{I}$ are the neutron and nuclear spins. The cross-section $\sigma_{\rm S}$ was obtained as $\sigma_{\rm S}=-0.26\pm 0.08~\rm{barn}$~\cite{okudaira2023_ImB}. Its theoretical expression is given by a function of $x$ and $y$ as~\cite{Gudkov2020}
\begin{equation}
\sigma_{\rm S}=0.079\left(-7x^2-2\sqrt{35}xy+\frac{2}{5}y^2\right).
\label{eq:ImB}
\end{equation}
Therefore, the solutions of $x$ and $y$ are obtained as, 
\begin{eqnarray}
(x, y)=&&(0.28\pm0.06,~0.96\pm0.02),\nonumber \\
&&(-0.96\pm0.02,~0.28\pm0.06), \nonumber\\
&&(-0.28\pm0.06,~-0.96\pm0.02), \nonumber\\
&&(0.96\pm0.02,~-0.28\pm0.06).
\label{xysolution}
\end{eqnarray}
Of the four possible solutions, the physical solution can be determined using other experimental results obtained by the angular correlation measurement of the $(n,\gamma)$ reactions.

The $\gamma$-ray angular correlations at the neutron-induced $p$-wave resonances arise from the interference between the partial amplitudes of the $s$- and $p$-wave resonances~\cite{fla85}. Therefore, the information of $x$ and $y$ can be extracted from the correlations of neutron spin $\sigma_{n}$, neutron momentum $k_n$, $\gamma$-ray spin $\sigma_{\gamma}$, and $\gamma$-ray momentum $k_{\gamma}$~\cite{fla85}. The angular correlation terms corresponding to $a_1$, $a_2$, and $a_3$, as described in Eq.~17 in Ref.~\cite{fla85}, for the $\gamma$-rays derived from the transition to the ground state of $^{140}$La were reported~\cite{okuda18,okd_2018Erratum,yama20,yama_errutum}. The measured values were obtained as a ratio to $a_0$, which represents the angular independent cross-section in the $(n,\gamma)$ reaction and is composed of the sum of both the $s$- and $p$-wave cross-sections, denoted as $a_{0s}$ and $a_{0p}$. 

The correlation terms $a_1$ and $a_3$, which correspond to the coefficient of the correlations $\vec{k}_n\cdot \vec{k}_\gamma$ and $(\vec{k}_n\cdot \vec{k}_\gamma)^2-1/3$, respectively, were measured through the angular distribution of $\gamma$-rays emitted from the $p$-wave resonance, which depends on $x$ and $y$. The equations of $x$ and $y$ related to $a_1/a_0$ and $a_3/a_0$ were obtained by comparing the experimental results and the theoretical expression of the angular correlations as~\cite{okuda18,okd_2018Erratum} 
\begin{eqnarray}
-0.409\pm0.024&=&0.30 x- 0.35 y\label{eq:ALH_a1},  \\
0.191\pm0.028&=&-0.20xy+0.033y^2\label{eq:ALH_a3},
\end{eqnarray}
respectively.

The correlation term $a_2$, which corresponds to the coefficient of the correlation $\vec{\sigma}_n\cdot(\vec{k}_n\times \vec{k}_\gamma)$, was measured through the transverse asymmetry of the $\gamma$-rays from the $p$-wave resonance, which also depends on $x$ and $y$~\cite{yama20,yama_errutum}. This transverse asymmetry, denoted as $A_{\rm{LR}}$, was obtained as $A_{\rm{LR}}=-0.60 \pm 0.19$~\cite{yama20,yama_errutum}. To compare with the theoretical expression due the contribution from $a_3$ in the denominator, $A_{\rm{LR}}$ should be multiplied by a factor of $(1-a_3/3a_{0p})$ ~\cite{yama20}. The theoretical expression, formulated as a function of $x$ and $y$ in Ref.\cite{fla85,yama20}, gives
\begin{equation}
A_{\rm{LR}}\left(1-\frac{a_3}{3a_{0p}}\right)=0.72x +0.42y.
\label{eq:ALR}
\end{equation}
Here, $a_3/3a_{0p}=(a_3/3a_{0})\times (a_0/a_{0p})=0.14\pm0.02$, where $a_3/a_{0}$ is obtained experimentally and $a_0/a_{0p}$ is calculated based on resonance parameters from Ref~\cite{okuda18}. Therefore, $A_{\rm{LR}}(1-a_3/3a_{0p})=-0.52\pm0.17$ is obtained. The analyses of $a_1$, $a_2$, and $a_3$ take into account the interference between the $p$-wave and neighboring $s$-waves listed in Ref.~\cite{EndoReso2023}.

\begin{figure}[]
	\centering
	\includegraphics[width=.43\textwidth]{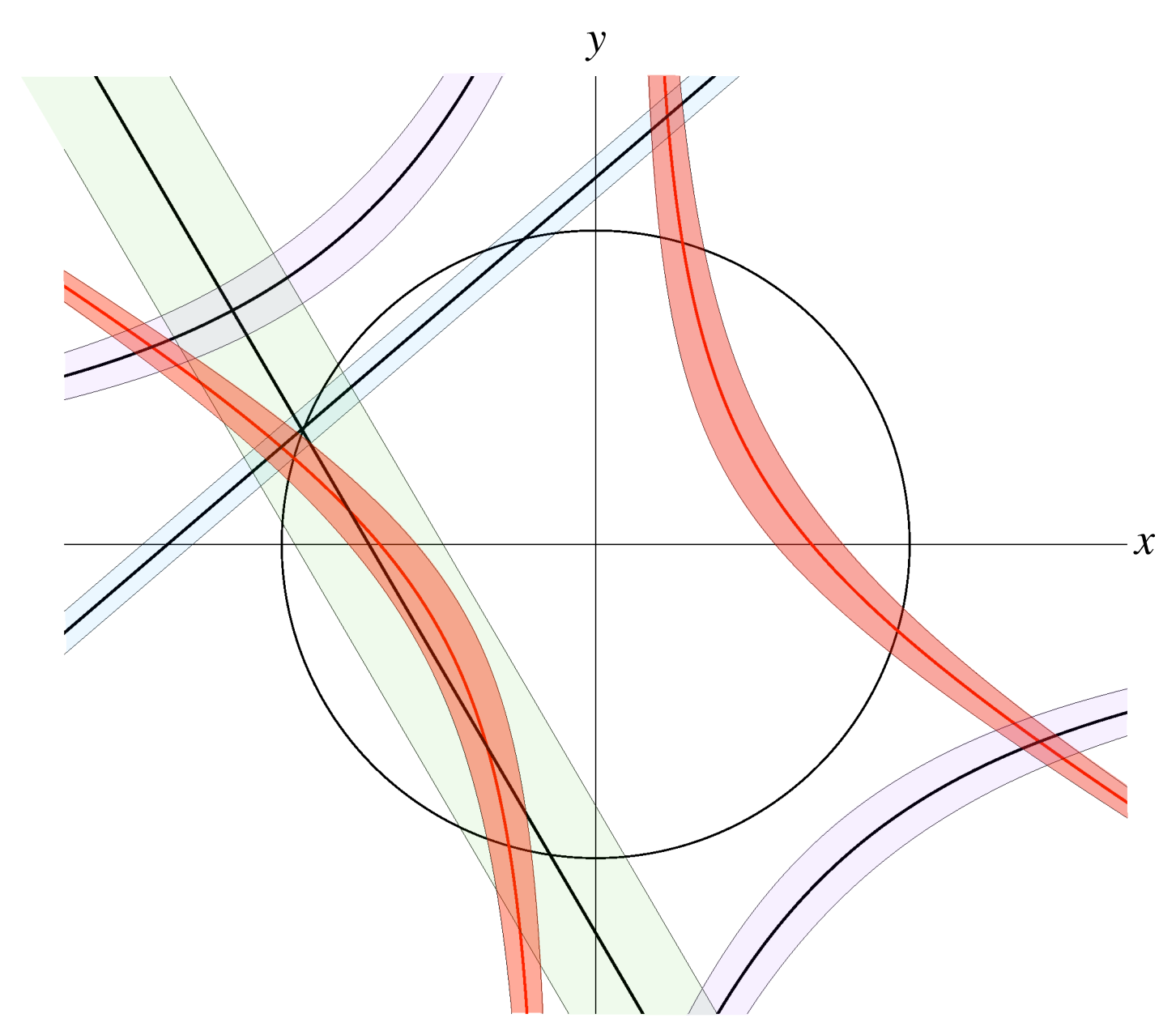}
	\caption{Visualization of the value of $\phi$ on the $(x,y)$ plane. The red, blue, green, and purple filled areas represent the neutron transmission, $a_1$, $a_2$, and $a_3$ results with a 1$\sigma$ region. The corresponding solid lines indicate the central values.}
	\label{fig_xyplane}
\end{figure}

Equations.~\ref{eq:ALH_a1}, ~\ref{eq:ALH_a3} and ~\ref{eq:ALR} are illustrated on the $(x,y)$ plane with Eq.~\ref{eq:ImB} as shown in Fig~\ref{fig_xyplane}. The unit circle expresses the relation $x^2+y^2=1$. The results of the $(n,\gamma)$ measurement were used to determine the physical solution from the four possible solutions in Eq.\ref{eq:ImB}. In the case where both spectroscopic parameters for $\gamma$-ray emission and neutron absorption are necessary, the neutron transmission measurement provides a more comprehensive understanding of the formation of compound states, in which large symmetry violation is expected. 

\begin{figure}[]
	\centering
 	\includegraphics[width=.47\textwidth]{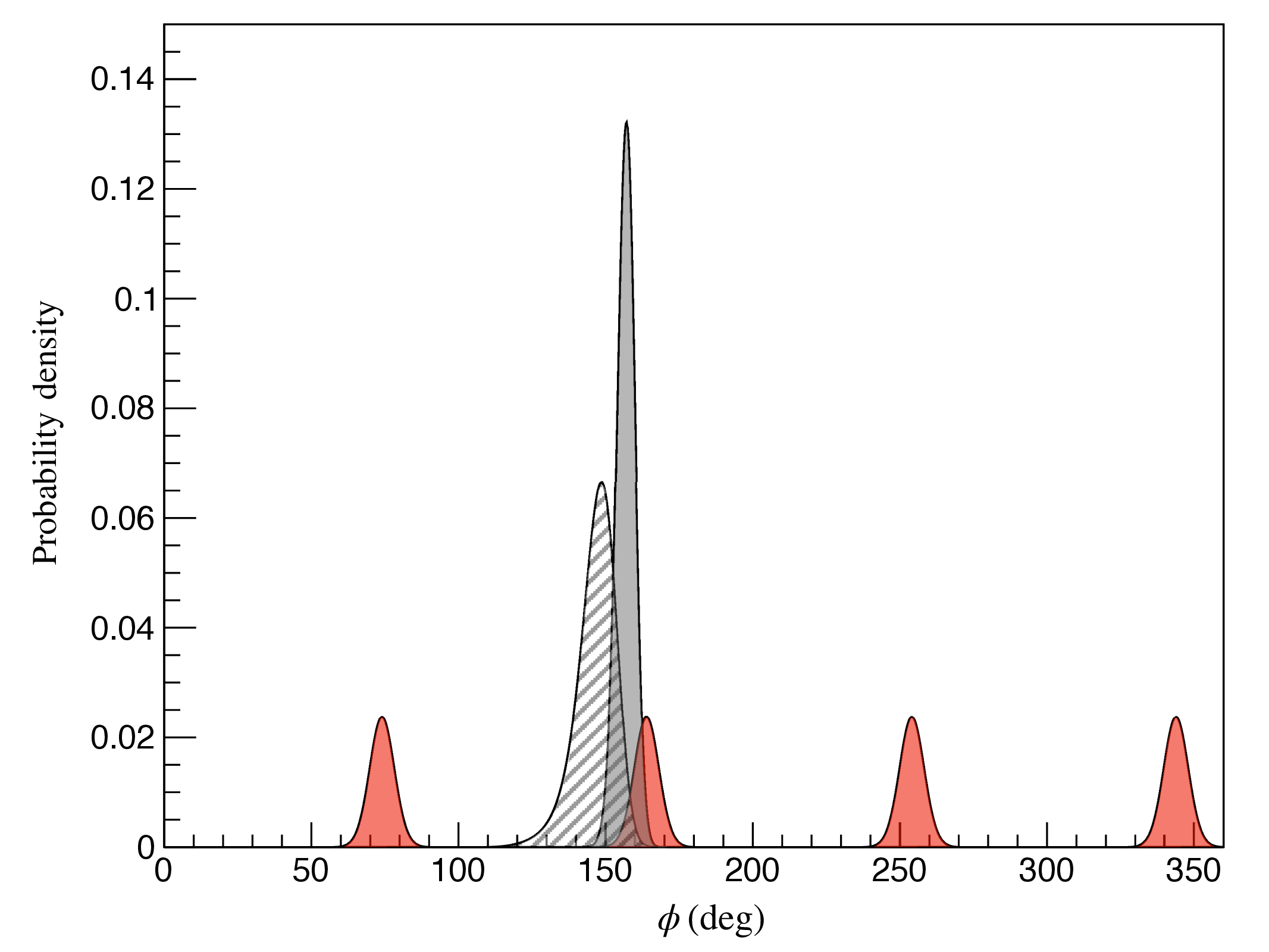}
	\caption{PDFs for $\phi$ on the unit circle. The red, shaded, and gray filled area represent the PDFs of Eq.\ref{eq:ImB}, the product of PDFs of Eqs.~\ref{eq:ALH_a1},~\ref{eq:ALH_a3}, and~\ref{eq:ALR}, and the product of PDFs of Eqs.~\ref{eq:ALH_a1},~\ref{eq:ALH_a3},~\ref{eq:ALR}, and \ref{eq:ImB}, respectively.}
	\label{fig_phi}
\end{figure}

Each result shown in Fig.~\ref{fig_xyplane} can be interpreted as a probability density function (PDF) on the unit circle. As depicted in Fig.\ref{fig_phi}, the PDFs suggest that the physical solution is in the second quadrant on the $(x,y)$ plane. The central value of Eq.~\ref{eq:ALH_a3} does not intersect the unit circle. Further study is necessary to identify the origin of this discrepancy. Thus the physical solution is obtained as $(x,y)=(-0.96\pm0.02,~0.28\pm0.06)$, which corresponds to $\phi=(164\pm4)^\circ$. Consequently, the spin-dependent factor in Eq.\ref{eq_kappa} is determined to be 
\begin{equation}
    \kappa=0.59\pm0.05.
\end{equation}
The $p$-wave resonance cross-section is calculated to be $3.06\pm0.09$~barn using the resonance parameters in Ref.~\cite{EndoReso2023}, and with the PV effect of $9.55\pm0.35$\% in Ref.~\cite{LANL91}, the TRIV cross-section in Eq.~\ref{eq_sigmaT} is
\begin{equation}
 \Delta\sigma_{\rm \not{T}\not{P}}=(0.17\pm0.02) \frac{W_{\rm T}}{W~{\rm}} {\rm (barn)}.
\end{equation}
The TRIV cross-section can be searched by means of measuring the transmission of low-energy polarized neutrons passing through a polarized target~\cite{STODOLSKY19865,Kabir1982,Kabir1988}. This gives the opportunity to improve the current limits of nucleon TRIV interactions. Moreover, this method is complementary to the ongoing measurements of electric dipole momenta. For example, theory predicts that $W_{\rm T}/W$ is sensitive to a linear combination of the isoscalar and isovector TRIV couplings in the $\pi$ meson exchange~\cite{Herczeg1987,Towner1994,Song2011_PV,Fadeev2019}. This method is a unique probe for the isoscalar coupling, since the isovector coupling is already tightly excluded by the measurement of the electric dipole moment of diamagnetic atoms ~\cite{HgEDM} while the neutron electric dipole moment~\cite{nEDM} is sensitive to the difference between the isoscalar and isovector TRIV couplings. 

This work was financially supported by JST SPRING, Grant Number JPMJSP2125, JSPS KAKENHI Grant Nos.17H02889, 19K21047, 20K14495, 23K13122. The author would like to take this opportunity to thank the “Interdisciplinary Frontier Next-Generation Researcher Program of the Tokai Higher Education and Research System.” C. J. Auton, J. G. Otero Munoz, and W. M. Snow acknowledge NSF grant PHY-2209481 and the Indiana University Center for Spacetime Symmetries. C. J. Auton acknowledges the Japan Society for the Promotion of Science. J. G. Otero Munoz acknowledges the NSF AGEP program, the GEM fellowship program, and the DOE SCGSR program. 
The design of the P-odd/T-odd measurement is in progress as a Project Category-II at RCNP, Osaka University and the J-PARC P99.
The authors would like to thank the staff members of beamlines 04 and 22 for the maintenance, MLF, and J-PARC for operating the accelerators and the neutron production target. 


\bibliography{PhiGlobalAnalysis}
\end{document}